\documentclass[prl,aps,twocolumn,amsmath,amssymb,showpacs]{revtex4}
\usepackage{graphicx}
\begin{document}

\title{Velocity Distributions of Granular Gases with Drag and with
Long-Range Interactions}

\author{K. Kohlstedt$^{1,2}$, A. Snezhko$^1$, M.V. Sapozhnikov$^{1,3}$,
I. S. Aranson$^{1}$, J. S. Olafsen$^{2}$, and E. Ben-Naim$^4$ }
\affiliation
{$^{1}$Argonne National Laboratory, 9700 S. Cass Avenue, Argonne, IL 60439 \\
$^{2}$Department of Physics and Astronomy, University of Kansas,
Lawrence, KS 66045\\
$^{3}$Institute for Physics of Microstructures, Russian Academy of
Sciences, GSP-105, Nizhny Novgorod 603600, Russia \\
$^4$ Theoretical Division and Center for Nonlinear Studies, Los
Alamos National Laboratory, Los Alamos, New Mexico 87545}

\begin{abstract}
We study velocity statistics of electrostatically driven granular
gases. For two different experiments: (i) non-magnetic particles in a
viscous fluid and (ii) magnetic particles in air, the velocity
distribution is non-Maxwellian, and its high-energy tail is
exponential, $P(v)\sim \exp\left(-|v|\right)$.  This behavior is
consistent with kinetic theory of driven dissipative particles. For
particles immersed in a fluid, viscous damping is responsible for the
exponential tail, while for magnetic particles, long-range
interactions cause the exponential tail. We conclude that velocity
statistics of dissipative gases are sensitive to the fluid environment
and to the form of the particle interaction.
\end{abstract}
\maketitle

Despite extensive recent studies, a fundamental understanding of
the dynamics of granular materials still poses a challenge for
physicists and engineers \cite{jnb,kadanoff,gennes}. Remarkably,
even dilute granular gases substantially differ from molecular
gases.  A series of recent experiments on granular gases, driven
either mechanically
\cite{menon,olafsen,kudrolli,losert,blair1,blair2,baxter} or
electrostatically \cite{ao}, reveals that the particle velocity
distribution significantly deviates from the  Maxwell-Boltzmann
distribution law. In particular, the high-energy tail of the
velocity distribution $P(v)$ is a stretched exponential
\begin{equation}
P(v)\sim \exp\left(-|v/v_0|^\xi\right)
\label{xi}
\end{equation}
with $v_0$ the typical velocity.  The exponent $\xi=3/2$ is observed
for certain vigorous driving experiments \cite{menon,ao}.
Non-Maxwellian velocity distributions were also observed in
experiments with a variety of geometries and driving conditions
\cite{olafsen,kudrolli,losert,blair1,blair2,baxter} and in numerical
simulations
\cite{grossman,carrillo,shattuck,soto,barrat,brey,vanzon}. Energy
dissipation is responsible for this behavior and this can be
understood using a simple model: a thermally driven gas of inelastic
hard spheres.  For high-energy particles, there is a balance between
loss due to inelastic collisions and gain due to the thermal driving.
For hard-core interactions, kinetic theory predicts (\ref{xi}) with
$\xi=3/2$, in agreement with vigorous shaking experiments
\cite{ernst}.  However, interactions between particles often do not
reduce to simple hard-core exclusion.

In this Letter, we study the effects of fluid environment and
particle interactions on electrostatically driven granular gases.
We perform experiments with particles immersed in a viscous fluid
and with magnetic particles in air subjected to an external
magnetic field. We find that the high-energy tail of the velocity
distribution is characterized by (\ref{xi}) but with the exponent
$\xi=1$.  We generalize the kinetic theory to situations with
viscous damping and with long-range interactions and find that the
experimental results are in-line with the kinetic theory
predictions. We conclude that velocity statistics in granular
gases depend sensitively on the environment and on the form of the
particle interaction.

\begin{figure}[ptb]
\vspace{-.4in}
\includegraphics*[width=0.43\textwidth,angle=-90]{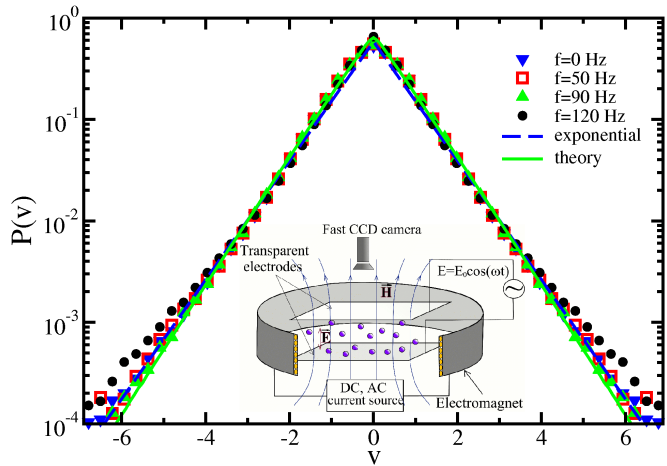}
\vspace{-.4in}
\caption{Velocity distribution functions for particles immersed in
fluid. The data are for 165 $\mu$m bronze particles immersed in
fluid (toluene) at frequencies $f=0, 50,90, 120$ Hz and applied
voltage $U=950$V. Corresponding rms velocities $v_{\rm rms}=$
1.68, 1.27, 0.75, 0.70 cm/s. Dashed line shows best fit to pure
exponential distribution $P(v)\sim\exp(-|v|/v_0)$  for $f=90$ Hz,
and solid line shows theoretical result from Eq.~(\ref{maxw1})
with $\eta=0.1$. Inset: Schematics of the experimental apparatus.}
\label{Fig1}
\vspace{-.2in}
\end{figure}

Our experimental setup is similar to that in
Ref.~\cite{blair,sapozhnikov1,sapozhnikov2}, see Inset to
Fig.~\ref{Fig1}.  The particles are placed between the plates of a
large capacitor that is energized by a constant (dc) or alternating
(ac) electric field $E=E_0 \cos(2 \pi f t)$. To provide optical access
to the cell, the capacitor plates were made of glass with a clear
conductive coating. We used $11\times11$ cm capacitor plates with a
spacing of 1.5 mm (big cell) or 4 cm diameter by 1.5 mm cell (small
cell). The particles are 165 $\mu$m diameter non-magnetic bronze
spheres or 90 $\mu$m magnetic nickel spheres. The field amplitude
$E_0$ varied from 0 to 10 kV/cm and the frequencies $f$ were between 0
and 120 Hz. The total number of particles in the cell is on the order
$10^6$. To control the magnetic interactions, the cell was placed
inside a large 30 cm electromagnetic coil capable of creating dc/ac
magnetic field $H$ up to 80 Oe. The cell can be filled with non-polar
dielectric fluid (toluene) to introduce viscous damping. The
electro-cell works as follows: conducting particles acquire a surface
charge when they are in contact with the capacitor plate. If the
magnitude of the electric field exceeds gravity, particles travel
upwards, recharge upon contact and then fall down. This process
repeats in a cyclic manner. By applying ac electric field and
adjusting its frequency $f$, one controls the vertical extent of
particles motion by effectively turning them back before they collide
with the upper plate, making the system effectively two-dimensional.

We extracted horizontal particle velocities using high-speed
videomicroscopy. Images were obtained in transmitted light at a rate
up to 2,000 frames per second from a camera mounted to a long focal
distance microscope. Particle positions were determined to sub-pixel
resolution. Inter-particle and particle-boundary collisions that
introduce sudden changes in momenta were filtered out in a manner
similar to Ref.~\cite{olafsen}. An ensemble average for each of the
velocity distributions was obtained from about $5 \cdot 10^6$ data
points.

We performed two sets of experiments: (i) electrostatically driven
non-magnetic particles in viscous fluid; (ii) electrostatically
driven magnetic particles in air  subjected to external magnetic
field. Some experiments were also performed with magnetic
particles in fluid. Although the origin of the particle
interaction is very different, both systems happen to show
somewhat similar behavior: exponential asymptotic velocity
statistics.  For the fluid system, the exponential behavior
results mostly from the dominant viscous drag.  However, the
effects of hydrodynamic dipole-dipole interaction between moving
particles in fluid  are  of certain importance:  the hydrodynamic
interaction between particles   become comparable with the viscous
drag if the particles are close enough or in contact
\cite{landau}. This interaction has  consequence for high velocity
tail, see discussion below. The ratio of viscous drag force $F_d$
to the gravity force $F_g$ at rms velocity in toluene is about
0.2-0.3 and in air is less then 0.007. Thus, viscous drag effects
are obviously dominant in toluene.  For the magnetic system the
exponential behavior is attributed to dominant long-range dipole
interaction since the air drag is negligible. Simple estimates
show that the magnetic dipole forces between particles dominate
gravity if the distance is smaller than 3 particles diameters.
Thus, due to remnant magnetization of the particles magnetic
interaction is dominant even for $H=0$.

Representative results for the fluid system are shown in
 Fig.~\ref{Fig1}.  Pure toluene was used in most experiments. Further
 experiments were performed using a toluene/polysterene mixture in
 order to control the viscosity of the solution, but no qualitative
 differences were found. Throughout this Letter, we analyze the
 distribution of the horizontal velocity components, $P(v)$, with
 $v\equiv v_x,v_y$. The velocity is normalized such that the
 root-mean-square (rms) velocity equals one, $\langle v^2\rangle=1$, and of
 course, the velocity distributions are symmetric, $P(v)=P(-v)$. As
 shown in Fig.~\ref{Fig1}, the velocity distributions are all notably
 different than the Maxwellian distribution. Moreover, the best fit to
 Eq.~(\ref{xi}) gives the value $\xi = 1$ in a wide range of
 parameters (driving amplitude and frequency). Remarkably, the
 velocity distributions for fluid-filled cells are different from
 those obtained for air-filled cells (where viscous drag is
 negligible) with $\xi=3/2$, other parameters the same \cite{ao}.

\begin{figure}[t]
\includegraphics*[width=0.45\textwidth]{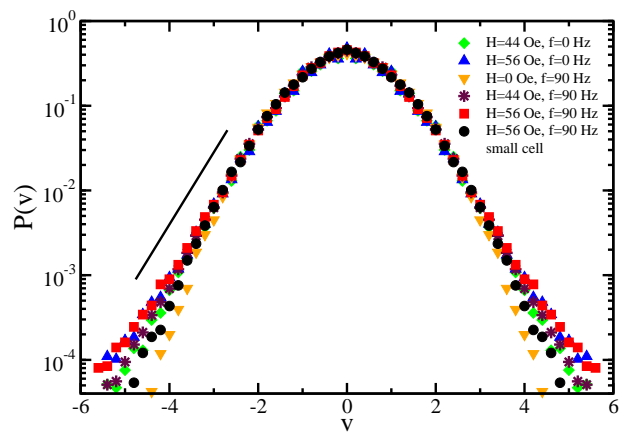}
\caption{Velocity distributions for magnetic particles. The data
 corresponds to 90 $\mu$m nickel spheres for dc and ac ($f=90$ Hz)
 electric field, $U=1000$V in $11\times 11$ cm cell, in dc magnetic
 field. Corresponding rms velocities $v_{\rm rms}$ for $f=0$ (dc) are
 1.9 cm/s ($H=44$ Oe) and 2.53 cm/s ($H=44$ Oe), and for $f=90$ Hz
 $v_{\rm rms}=3.0, 2.12, 2.34$ for $H=0,44,56$ Oe.  Data for a small
 $3$ cm cell is also shown for comparison. Exponential decay is
 indicated by straight line as a reference.}
\label{Fig2}
\vspace{-.2in}
\end{figure}

Experiments with magnetic interactions were performed using nickel
magnetic microparticles with an average size of about 90$\mu$m
(Alfa Aesar Company). The magnetic moment per particle at 80 Oe is
$~1\cdot10^{-5}$ emu; the saturated magnetic moment is
$2\cdot10^{-4}$ emu per particle and the saturation field is about
4 kOe.  A vertically oriented external magnetic field was applied
in order to control the magnetic interactions: since the particles
are multi-domain, and nickel is a soft magnetic material, the
applied field can effectively increase the particle's magnetic
moment. The corresponding velocity distributions are shown in
Fig.~\ref{Fig2}. The magnetic field systematically widens the
velocity distribution and it enhances the exponential asymptotic
decay of $P(v)$ at the high velocities. This observation is
consistent with the fact that the applied magnetic field enhances
the dipole-dipole magnetic interaction due to the magnetization of
the particles. Comparing the fluid and the magnetic systems, we
note that the velocity distributions have different cores, but the
tails are exponential in both cases.

In the course of the measurements we noticed that finite size effects
have a strong influence on the tail of $P(v)$. We performed a number
of measurements using a 4 cm diameter cell, and about 3,000
particles. At the same conditions, the velocity distribution in the
larger cell has a more pronounced exponential tail (see
Fig.~\ref{Fig2}).

We also carried out several experiments combining features of the two
systems using magnetic particles in toluene solution, and obtained
results consistent with the rest of our observations: non-Maxwellian
velocity distributions with an exponential tail.  Compared with pure
magnetic interactions results, the range of exponential behavior for
the tail becomes even more extended.

We now compare the experimental results with the predictions of
kinetic theory, the standard framework for describing granular
gases. This requires generalization of (\ref{xi}) to situations
with long-range interactions.  For particles interacting via the
potential \hbox{$U(r)\sim r^{-\sigma}$}, $r$ is the interparticle
distance, the collision rate $K$ grows algebraically with the
normal component of velocity difference $\Delta v$, $K\propto
(\Delta v)^\lambda$ with $\lambda=1-2\frac{d-1}{\sigma}$, and $d$
is the dimension of space \cite{rd}. Hard-spheres, $\lambda=1$,
model granular particles with hard-core ($\sigma\equiv \infty$)
interactions while Maxwell molecules \cite{mhe},
\hbox{$\lambda=0$}, model granular particles with a specific
dipole interaction. In two-dimensions, relevant to our
experiments, the collision rate effectively becomes independent on
the relative velocity when $\sigma=2$.

First consider magnetic particles. To analyze the high-energy tail, we
make the standard assumption that forcing is thermal. For energetic
particles, gains due to collisions are negligible and losses due to
collisions are balanced by the forcing. The high-energy tail of the
velocity distribution is governed by this balance,
\hbox{$d^2P(v)/d^2v\propto K(v)P(v)\propto v^\lambda P(v)$}
\cite{bk}. Consequently, the velocity distribution decays as a
stretched exponential (\ref{xi}) with $\xi=1+\lambda/2$. In general,
the velocity statistics are non-Maxwellian and the tails are
over-populated with respect to a Maxwellian distribution. The upper
limit, $\xi=3/2$, is realized for hard-spheres, and pure exponential
behavior, $\xi=1$, occurs for Maxwell molecules. For magnetic dipole
interactions one has $\sigma=2$. In this case, in two dimensions the
kinetic theory predicts a simple exponential tail, $\xi=1$.

In the following, we consider a representative data set for the fluid
system (120 Hz) and for the magnetic system (44 Oe, 90 Hz).  We
display a single set because variations in $P(v)$ among the different
experimental conditions are relatively small: the kurtosis
$\kappa\equiv \langle v^4\rangle/\langle v^2\rangle^2$ varies by less
than 5\% for the different data sets.

\begin{figure}[ptb]
\includegraphics*[width=0.45\textwidth]{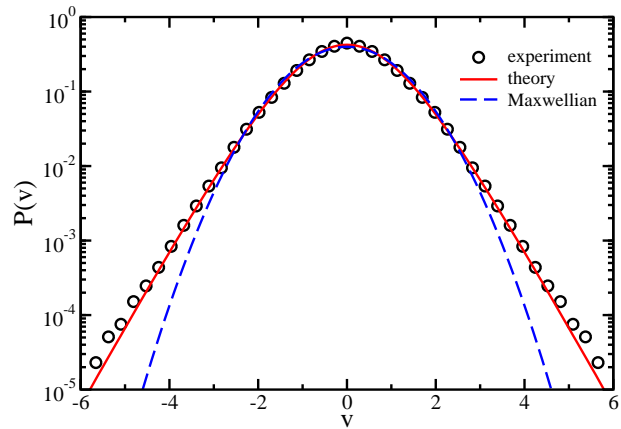}
\caption{The experimental distribution for magnetic particles vs
the theory for forced Maxwell molecules. Crossover to exponential
law occurs for $|v|>2$.} 
\label{Fig4}
\end{figure}

For the fluid system, the velocity distribution is very close to a
pure exponential, as shown in Fig. \ref{Fig1}, and furthermore,
the kurtosis $\kappa= 6.2\pm 0.2$ is within 3\% of the value
corresponding to a pure exponential distribution $\kappa=6$.
Viscous damping is responsible for this behavior and the nearly
exponential distribution is consistent with the damping process
$v\to \eta v$ suggested by van Zon et al \cite{vanzon}.  This
mimics viscous damping because $v_n=v_0\eta^n$ \cite {vanzon} with
$n$ the number of damping events and $n$ growing linearly with time.
The damping rate is set by the frequency of collisions with the
plates, but in the theory, it can be set to one without loss of
generality. When the viscous dissipation dominates over the
collisional dissipation, the kinetic theory is modified as follows
\begin{equation}
D\partial^2_vP(v)+ \eta^{-1} P\left(v/\eta\right)-P(v)=0
\label{maxw1}
\end{equation}
where the last two terms represent gain and loss due to drag and $D$
is diffusion coefficient. At large velocities, the gain term is
negligible and consequently, the tail is exponential $P(v)\sim
\exp\left(-v/\sqrt{D}\right)$. Eq. (\ref{maxw1}) can be solved
analytically, and there is a family of velocity distributions
characterized by one parameter $\eta$.  When $\eta\to 0$, the gain
term in (\ref{maxw1}) is negligible and the distribution is purely
exponential. This is reflected by the kurtosis
$\kappa=\frac{6}{1+\eta^2}$. For strong damping, $\eta\to 0$, the
value corresponding to a pure exponential distribution is realized,
$\kappa\to 6$.

A Discrete Simulation Monte Carlo methods was used to solve the
Boltzmann equation for Maxwell molecules. In the simulations pairs of
randomly chosen particles collide according to the inelastic collision
rule $({\bf v}_1,{\bf v}_2)\to({\bf v}_1^\prime,{\bf v}_2^\prime)$
with $({\bf v}_1^\prime-{\bf v_2}^\prime)\cdot \hat {\bf n}=-\alpha
({\bf v}_1-{\bf v_2})\cdot \hat {\bf n}$ and ${\bf v}_1^\prime+{\bf
v_2}^\prime={\bf v}_1+{\bf v_2}$ with $\alpha$ the restitution
coefficient and $\hat{\bf n}$ the impact direction. In addition,
particles are thermally forced $dv/dt=\zeta$ with $\zeta$ a white
noise.  Also, damping $v\to \eta v$ with unit rate models the fluid
effect.  The simulation results represent an average over $10^2$ runs
in a system with $N=10^7$ particles.

When viscous damping dominates over collisional dissipation, the
distribution is nearly exponential, see Fig.~\ref{Fig1}, and in a
very good agreement with the experiments. We note that even though
the drag term dominates over the collision terms, the experimental
results suggest that the collision rate is velocity independent at
least at large velocities, $\lambda=0$. If this were not the case,
the collisional loss term would dominate at some very large
velocity and there would be a cross over to $\xi>1$ \cite{foot1}.
But no such crossover is observed experimentally. We conclude that
the results of kinetic theory of forced Maxwell molecules with
strong viscous damping are consistent with the experimental
results for all velocities.

For magnetic interactions, there is an excellent agreement between
the experiment and the theory of thermally forced Maxwell
molecules for which the collision rate is completely independent
on the relative velocity (see Fig. \ref{Fig4}). The kurtosis,
$\kappa = 3.6\pm 0.1$, falls within 2\% of the analytically known
value $\kappa=3+\frac{18}{33}\cong 3.55$ obtained from $\kappa
=3+\frac{18p^2(1-p)}{(d+2)(1+p)-3(1-p)(1+p^2)}$ where
$p=(1-\alpha)/2$ \cite{bk}.  We note that there are no fitting
parameters. The restitution coefficient $\alpha$ was set to zero
because particle collisions in the experiments are strongly
inelastic and because as long as the dissipation is strong, there
is only a weak dependence on $\alpha$.  Even though the tail of
the distribution is close to a simple exponential, its core is
approximately Maxwellian, as reflected by the kurtosis that is
much closer to the pure Maxwellian value of 3 than the pure
exponential value of 6. We conclude that for magnetic particles
the collision rate becomes practically independent on the relative
velocity, and conversely, that they are accurately modeled by
Maxwell molecules.

In summary, our main result is that velocity statistics of forced
granular gases depend sensitively on the fluid environment and on the
nature of the inter-particle interactions. The two sets of experiments
can be universally described by a specific version of the kinetic
theory, Maxwell molecules, with a velocity independent collision
rate. Whereas dipole interactions are ubiquitous for the magnetic
system, our studies indicate that the Maxwell model is the only way to
interpret the fluid experiments.  If hard sphere were used, the
collision rate will grow linearly with the velocity and $\xi=3/2$
stretched exponential tail will prevail at large velocities. No such
crossover is observed and this is a strong evidence that the collision
rate is velocity independent. Thus, although the core behavior is
dominated by the damping, the tail behavior indicates that long range
interactions play a role. In view of this, the two experiments are
complementary.

We comment that it is difficult to experimentally validate that
the driving is thermal in nature. However, the consistent
agreement between the experiments and theory supports this
widely-used modeling assumption. In addition, the excellent
quantitative agreement between the magnetic particles experiments
and the Maxwell molecules kinetic theory suggests that magnetic
particles are an ideal experimental probe for the predictions of
this analytically tractable theory, including in particular, the
transport coefficients \cite{as}. We also propose that stretched
exponential velocity distributions may be generic for dissipative
gases with competing interactions, and may possibly be relevant
for vastly different systems, such as dusty plasmas, colloids, and
even star clusters, where long-range interactions (e.g due to
gravity) are mediated by short-range collisions. 

We thank R.~Brito and M.~Ernst for useful discussions and acknowledge
support from the US DOE, Office of Science, contracts W-31-109-ENG-38
(IA,AS) and W-7405-ENG-36 (EBN), the General Research Fund at the
University of Kansas and Kansas NASA EPSCoR (JSO).

\end{document}